# Strategy for enhanced thermoelectric performance of $Bi_2S_3$ nanorods by Bi nanoinclusions

Tarachand,[1] Gunadhor Singh Okram,[*,1] Binoy Krishna De,[1] Siddhartha Dam,[2] Shamima Hussain,[2] Vasant Sathe,[1] Uday Deshpande,[1] and Archana Lakhani[1]

[1]UGC-DAE Consortium for Scientific Research, University Campus, Khandwa Road, Indore-452001, India.
[2]UGC-DAE Consortium for Scientific Research, Kalpakkam Node, Kokilamedu-603104, India.

**ABSTRACT:** This is the first report on the enhanced thermoelectric (TE) properties of novel $Bi_2S_3$-Bi nanocomposites synthesized using a one-step polyol method at different reaction temperatures ($T_{Re}$) and time. They are well-characterized as nanorod-composites, coexistent with orthorhombic $Bi_2S_3$ and rhombohedral Bi phases together in which the latter coats the former forming $Bi_2S_3$-Bi core-shell type structures along with independent Bi nanoparticles (NPs). There is a very significant observation of systematic reduction in electrical resistivity $\rho$ with reaction temperature and time duration increase, revealing a promising approach for reduction of $\rho$ in this highly resistive $Bi_2S_3$ and hence resolving the earlier obstacles for its thermoelectric application potentials for the past few decades. Most astonishingly, TE power factor at 300 K of highest Bi content nanocomposite pellet, made at 27 ºC using ~900 MPa pressure, is 3 orders of magnitude greater than that of hot-pressed $Bi_2S_3$, or even 23% better than that of spark plasma-sintered core-shell $Bi_2S_3$@Bi sample reported earlier (Tarachand *et al.* Nano Res. 2016, 9, 3291; Ge *et al.* ACS Appl. Mater. Interfaces 2017, 9, 4828). Considering the probable greatly reduced thermal conductivity due to their complex nanostructures, the significantly improved TE performance potential near 300 K is highly anticipated for these toxic- and rare earth element-free TE nanocomposites, making the present synthesis method as a pioneering approach for developing enhanced thermoelectric properties of $Bi_2S_3$-based materials without using extra sintering steps.

## 1. Introduction

Energy scarcity in the world due to the depletion of fossil fuels is attracting special attention towards thermoelectric (TE) materials that are environmental-friendly and produce waste heat-generated electrical energy without pollutions. The efficiency of TE materials is determined from the thermoelectric figure of merit defined as $ZT = \alpha^2\sigma T/\kappa$, where $\alpha$, T, $\sigma$ and $\kappa$ are Seebeck coefficient, absolute temperature, electrical and thermal conductivity, respectively.[1-5] Main challenge in this field is to obtain high TE power factor ($\alpha^2\sigma$) value while maintaining low $\kappa$. The understanding in recent years on nanostructuring as a boon to enhance ZT through significant suppression of $\kappa$[3,4,6] has met with roadblock of lower limit of $\kappa$ beyond which it cannot be further reduced. In addition, $\alpha$ can be enhanced by resonant doping or low energy charge carriers filtering. However, doping in nanomaterials is a very difficult process[1] because it needs sufficiently high temperature reaction at which nanomaterial can be transformed into bulk material. Hence, the band engineering with the introduction of nanoparticles (NPs) in nanostructured materials is considered to be the best way through which one can enhance both $\alpha$ and $\sigma/\kappa$.[2-4,7,8] Sumithra *et al.*[2] investigated the effect of Bi (semimetal) nanoinclusion in nanostructured $Bi_2Te_3$ and found 100% enhancement in ZT for composites with 5%–7% Bi NPs. They found that presence of Bi NPs in $Bi_2Te_3$ not only reduced $\kappa$ (due to phonon scattering) but also enhance $\alpha$ (due to low energy electron filtering at the interface). Similarly, Tarachand *et al.*[3] reported 34% improvement in the TE power factor at 300 K for Cu-doped $Bi_2S_3$ as compared to pure $Bi_2S_3$ hot pressed at 480 ºC due to formation of $Bi_2S_3$@$Bi_{4.8}Cu_{2.94}S_9$ nanocomposites.

However, the relatively toxic and high cost of the presently available commercially efficient TE materials face two main challenges. In this, $Bi_2S_3$ as a potential candidate attracts special attention due to high $\alpha$ and low $\kappa$.[3,4,6] It belongs to the family of most promising TE materials $Bi_2X_3$ (X = Te, Se and S). It is a direct band gap semiconductor and crystallizes into orthorhombic phase at room temperature. In spite of its high electrical resistivity ($\rho$), many reports have already been available on the attempts to reduce $\rho$ by creating sulfur deficiencies through heat treatment as well as introduction of conducting nanoparticles in the resistive host $Bi_2S_3$ nanomaterials.[3,4] For this, Bi NPs looks suitable due to its several attractive features such as low melting point (271 °C), highly anisotropic Fermi surface, small carrier effective masses, low carrier density, long mean free path of ~1 mm that renders very low thermal conductivity[9], a semimetal with a narrow-band overlap (38 meV) and rhombohedral crystal structure.[10] The Fermi surface of the bulk Bi consists of a single hole pocket at the T point and three highly elongated electron pockets at the L point.[11,12] Theoretically, its ZT is ~ 6 at 77 K in 5 nm bismuth nanowire.[5] Therefore, combination of relatively conducting Bi with highly resistive $Bi_2S_3$ may produce highly efficient TE materials.

Recently, Ge *et al.*[4] reported the formation of $Bi_2S_3$@Bi core-shell nanowires with controlled Bi shell thickness using variation in reaction time in the presence of hydrazine. Spark plasma (SP) sintered pellets of the powder at 673 K for 1 h resulted in the highest ZT value of 0.36 at 623 K. However, the lattice thermal conductivity ($\kappa_L$) was increased as a result of converting Bi into its bulk since Bi got melted during SP sintering at 673 K due to it being above its melting point (270 °C) as extruded from cracks of the graphite die. This led also to non-uniform distribution of Bi in the composites. Consequently, $\sigma$ ($1/\rho$) and $\kappa$ for $Bi_2S_3$@Bi containing higher Bi were increased, thereby its ZT was low. Thus, SP sintering at higher temperature turns out to be inappropriate to improve the efficiency for $Bi_2S_3$@Bi nanocomposites.

In this direction, there are numerous bottom up strategies (solvothermal,[2] polyol,[3] hydrothermal[4]) already developed for synthesis of such nanocomposites. Herein, we took the advantage of a cheaper and environmental-friendly simple polyol method to prepare $Bi_2S_3$-Bi nanocomposites with optimization process of reaction temperature and time. Just varying reaction temperature and time lead to exceptionally remarkable colossal reduction in resistivity for the first time, which will open a promising strategy for improving ZT value in $Bi_2S_3$-based TE materials without using any extra sintering process like SP sintering.

## 2. Experimental section

**Method** For synthesis of $Bi_2S_3$-Bi nanocomposites, initially 2 mmol of $Bi(NO_3)_3 \cdot 5H_2O$ ($\geq$ 98 %, Merck) and 3.5 mmol of thiourea ($\geq$ 98 %, Merck) were dissolved in 50 mL di-ethylene glycol (DEG) in a three-neck round-bottomed flask by a using magnetic stirrer. The resultant yellowish solution was heated at the rate of 6 °C/min, under continuous Ar-gas flow to avoid any oxidation, and after 10 min, colour of the solution turned into milky-white, then black at 100 °C. The reaction temperature ($T_{Re}$) was further increased to 200 °C, 220 °C and 240 °C for 2 h each, which produced the black precipitates of $Bi_2S_3$ nanoparticles (NPs). They are named as BS200, BS220 and BS240, respectively. When $T_{Re}$ touches 245 °C, the boiling off of DEG started, that leads to reduction of its quantity gradually with time which causes in turn more volatilization of sulphur. For this, $T_{Re}$ is initially maintained $\leq$200 °C for nearly 2 h then increased up to 245 °C and maintained it for 1.5 h, 3 h and 4.5 h after which 50 mL reaction solution has got reduced to 44 mL, 40 mL and 35 mL, respectively; they were named as BS245, BS245a and BS245b, respectively. After this, for synthesis of single phase Bi NPs, we just removed thiourea from above reaction and heated at 240 °C directly for 2 h. Washing steps used in earlier report[7] were followed in washing that gives black precipitates of NPs. The precipitates were dried at 60 °C in vacuum for 1 h. Resulting powders are directly used for further characterizations.

**Experimental techniques** Powder X-ray diffraction was done using Bruker D8 Advance X-ray Diffractometer with Cu $K_\alpha$ radiation (1.54 Å) in the (2θ) angle range of 10° to 70°. The energy dispersive analysis of X-ray (EDX) measurements were carried out using a JEOL JSM 5600 scanning electron microscope equipped with EDX. Morphological study with EDX analysis was performed using field emission scanning electron microscopy (FESEM) using a Carl Zeiss AURIGA FIBSEM in secondary emissions mode. X-ray photoelectron spectroscopy (XPS) measurements were performed on an X-ray photoelectron spectroscope (SPECS, Germany) using Al $K_\alpha$ radiation with an anode voltage of 13 kV and an emission current of 22.35 mA. A survey scan spectrum was collected with an energy of 40 eV and high-resolution spectra were collected with an energy of 30 eV to know the valence states of the constituent elements. Zeta potential and hydrodynamic diameter were measured in de-ionized water with pH 7 at 27 °C using a zeta/particle size analyser NanoPlus-3. Raman spectra were recorded at room temperature using a Jobin Yvon Horiba LABRAM-HR Visible instrument equipped with an Ar ion laser of wavelength 473 nm. Thermopower ($\alpha$) measurements using a load-based differential direct-current techniques,[13] and resistivity measurements using two and four probe methods[14] in temperature range of 5 - 325 K were performed in an especially designed commercially available Dewar in a home-made setup. Hall measurements using a five probe method were performed on a 9 T AC Transport PPMS (Quantum Design) system.

## 3. Results and discussion
### 3.1 X-ray Diffraction and energy dispersive analysis of X-ray

Systematic optimizations of reaction conditions such as quantity of initial precursors, reaction temperature and time are essential for making any good quality samples. We reported earlier the synthesis of single phase $Bi_2S_3$ nanorods at 180 °C by polyol method and found that below this, intensity of characteristic peaks is suppressed and produced low yield due to incomplete reaction.[3] In the case of sample preparation above 180 °C, some extra impurity peaks were observed which were not identified in the earlier report.[3] Herein, we identified those extra peaks (Fig. 1) and found that all the extra peaks are well-matched with those of rhombohedral bismuth (Bi) (JCPDS 851329),[15] except one low intensity peak centered at 30.56°. It was suggested that this extra peak (at 30.56°) is due to formation of metal-sulfite (Bi-$SO_x$) and can be removed by increasing the quantity of sulfur source.[3] Therefore, we increased the quantity of sulfur source slightly (3.5 mmol) here than that of nominally required (3 mmol) one. The powder XRD patterns of effect of reaction temperature of the samples synthesized at 200 °C (BS200), 220 °C (BS220), 240 °C (BS240), then at $T_{Re}$ = 245 °C at the reaction time of 1.5 h (BS245), 3 h (BS245a) and 4.5 h (BS245b) are shown in Fig. 1 (a). All the peaks are appropriately matched with reference card JCPDS#652431 of $Bi_2S_3$ and JCPDS 851329 of Bi. Notably, the intensity of peaks belonging to Bi-phase is seen to be enhanced with increasing $T_{Re}$ that corresponds to enhanced mass density of the composites (Table S1). The average particle size determined from Scherrer formula is in the range of 21.6 nm to 41.3 nm for $Bi_2S_3$ and 44.7 nm to 58.7 nm for Bi NPs (Table S1). Rietveld refinement[16] of powder XRD of these nanocomposites confirmed the existence of heterostructured dual phase of orthorhombic $Bi_2S_3$ and rhombohedral Bi. For this, initial refinement parameters were taken from ref. 3 for $Bi_2S_3$ and ref. 15 for Bi phases, and refined parameters are tabulated in Table 1. Figure 1 (b) is the Rietveld fitting graph of BS240 and those for others are shown in Fig. S1 and that for Bi is shown in Fig. 1 (c). First sample, BS200, contains only 0.65% of Bi phase with 99.35 % $Bi_2S_3$ (Table 1). The fraction of Bi-phase systematically rises with increasing $T_{Re}$ (Table 1). However, even though an excess quantity of sulfur source (3.5 mmol), as compared to that of nominally required quantity (3 mmol), has been taken, due to volatile nature of sulfur, a sulfur-deficient sample is formed, in favor of the formation of Bi NPs in addition to $Bi_2S_3$. As a result, the composites tend to be Bi-riched composites with increasing $T_{Re}$ and reaction time (Tables 1 and S1).

The calculated average crystallite size ($L_{av}$) using Scherrer formula is shown in Fig. 2 (a) and Table S1. The higher $T_{Re}$ and reaction time increase the growth rate of crystals, well reflected in the calculated L of both phases ($Bi_2S_3$ and Bi). The average crystallite size of $Bi_2S_3$ phase ($L_{Bi2S3}$) in the nanocomposites increases up to 41.1 nm (for BS240) with increasing $T_{Re}$. Then, $L_{Bi2S3}$ decreases with further increase in $T_{Re}$ and reaction time, due to most probably volatilization of sulfur (Fig. 2 (a)). Whereas the average L of Bi-phase ($L_{Bi}$) in nanocomposites continuously increases with increasing $T_{Re}$ and time (Fig. 2 (a)), that is well-corroborated with increase in Bi-phase fraction obtained from Rietveld refinement (Table 1). This seems to suggest the formation of core-shell structure of $Bi_2S_3$-Bi along with Bi NPs in which core or shell layer thickness appears to vary with $T_{Re}$ and time. This has been indirectly confirmed from the very significant decrease in their electrical resistivity with increase in conducting Bi phase coated over highly resistive phase of $Bi_2S_3$, as discussed later.

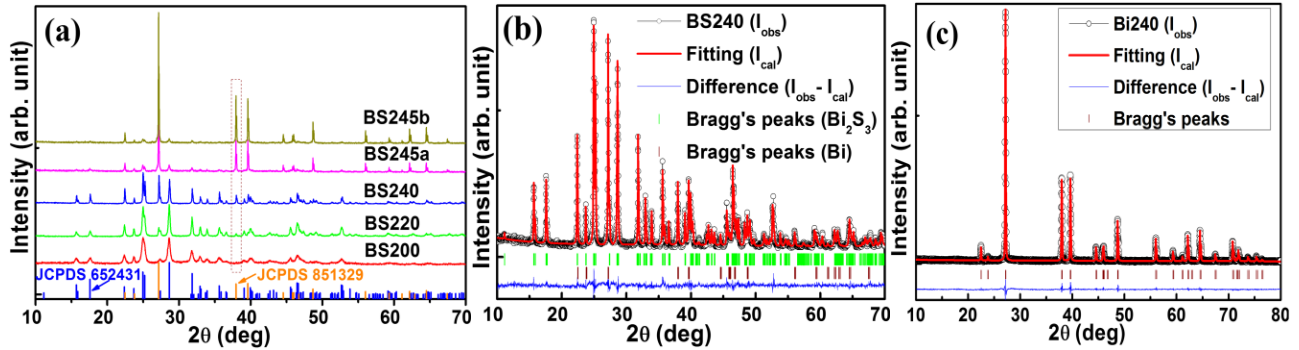

**Figure 1** (a) Powder XRD of $Bi_2S_3$-Bi nanocomposites synthesized by polyol method with different reaction temperature, and Rietveld refinement of (b) $Bi_2S_3$-Bi nanocomposites (BS240) and (c) Bi nanoparticles (Bi240) prepared at 240 ºC.

**TABLE 1.** The obtained crystallographic parameters (lattice constants (a, b, c), volume (V), mass density ($\rho_v$) and phase fraction (f)) after Rietveld refinement of powder XRD of nanocomposites consisting orthorhombic (**O**) $Bi_2S_3$ and rhombohedral (**R**) bismuth (Bi) phases; standard errors are shown in parentheses.

| Parameters | | BS200 | BS220 | BS24 | BS245 | BS245a | BS245b | Bi |
|---|---|---|---|---|---|---|---|---|
| a (Å) | O | 11.298(3) | 11.302(1) | 11.298(1) | 11.295(3) | 11.298(2) | 11.295(3) | |
| | R | 4.565(2) | 4.546 (1) | 4.545(1) | 4.546(1) | 4.545(1) | 4.546(1) | 4.546(1) |
| b (Å) | O | 3.981(1) | 3.981(1) | 3.981(1) | 3.982(1) | 3.982(1) | 3.982(1) | |
| c (Å) | O | 11.149(3) | 11.149(1) | 11.147(1) | 11.144(3) | 11.147(2) | 11.144(3) | |
| | R | 11.824 | 11.863(3) | 11.859(1) | 11.860(1) | 11.860(1) | 11.860(1) | 11.862(1) |
| V (Å$^3$) | O | 501.437 | 501.665 | 501.386 | 501.247 | 501.405 | 501.247 | |
| | R | 213.380 | 212.300 | 212.110 | 212.206 | 212.195 | 212.206 | 212.32 |
| $\rho_v$ (g/cm$^3$) | O | 6.811 | 6.822 | 6.811 | 6.813 | 6.811 | 6.813 | |
| | R | 9.758 | 9.808 | 9.816 | 9.812 | 9.813 | 9.812 | 9.807 |
| f (%) | O | 99.35 | 95.69 | 93.19 | 41.99 | 34.57 | 16.48 | |
| | R | 00.65 | 04.31 | 06.8 | 58.01 | 65.43 | 83.52 | 100 |

For further analysis, an EDX measurement is performed which gives the elemental compositions of $Bi_2S_3$-Bi nanocomposites (Fig. 2 (b-d)) and atomic percentage constituent elements (expected and experimentally obtained) are tabulated in Table S2. The ratio of atomic % of Bi:S viz., 0.685, 0.726 and 2.95 for BS200, BS220 and BS245a, respectively, is significantly greater than that of pure $Bi_2S_3$ (0.667) synthesized at $T_{Re}$ = 180 ºC.[3] These results are in accordance with systematic increase in Bi-phase fraction with increasing $T_{Re}$ and time observed in XRD analysis (Fig. 1 and Table 1).

**3.2 Field Emission Scanning Electron Microscopy (FESEM)**

Figure 3 shows the FESEM images of $Bi_2S_3$-Bi nanocomposites synthesized at $T_{Re}$ = 240 ºC and 245 ºC. For BS240 sample the most of particles are in rod shaped (Fig. 3(a) and (b)) with diameter in the range of 35-190 nm. The average diameter of nanorods (NRs) is 55 nm and its length ranges from 40 nm to 2 μm. This formation of NRs in polyol method of synthesis is consistent with earlier report synthesized at 180 ºC.[3] Due to high $T_{Re}$, the diameters of these NRs are larger than that of $Bi_2S_3$ NRs synthesized at 180 ºC.[3] Along with these NRs, some chunk-like particles also are observed with thickness ~80 nm and width in the range of few hundreds of nm. In FESEM image of BS245a (Fig. 3(c)) aspect ratio (length to diameter) of NRs is reduced and number of chunks/disks like particles has increased. Due to the boil off of DEG at high reaction temperature (245 ºC), sulfur atoms are more volatilized thereby only longer NRs (of $Bi_2S_3$) with bigger diameter have been survived along with formation of rectangular or spherical or chunk-like nanoparticles (Fig. 3(c)).

In order to better understand the elemental composition of nanorods, we performed EDX measurements by focusing electron beam onto very small portion of individual nanoparticles. The obtained Bi:S atomic % ratio are 81.88:18.12, 58.25:41.75 and 61.42:38.58 for chunk (Fig. 4(a)), nanorod (Fig. 4(b)) and nearly spherical particle (Fig. S2), respectively for BS245a. Since only Bi semimetal and $Bi_2S_3$ are formed as per XRD results, this means that atomic % of Bi:S::2:3 for $Bi_2S_3$ may be

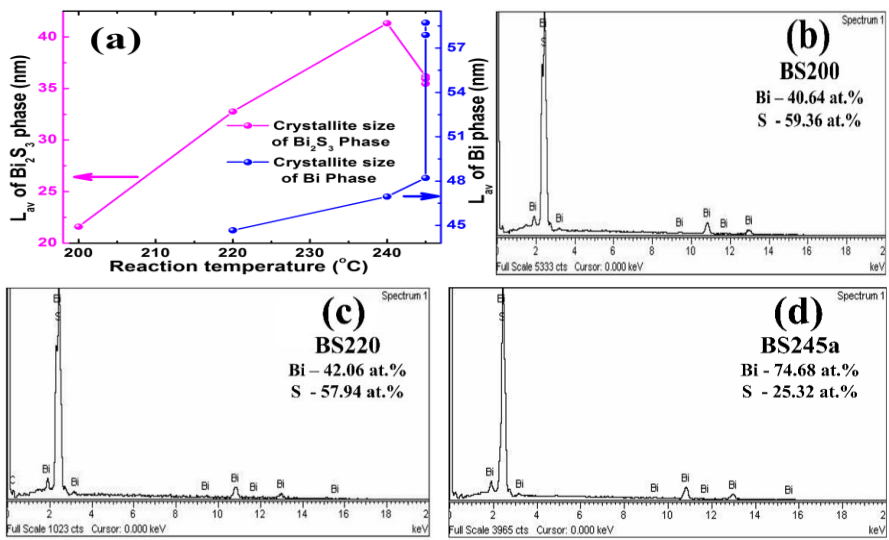

**Figure 2** (a) Calculated crystallite size for both phases, $Bi_2S_3$ (left) and Bi (right), of the nanocomposites synthesized at different reaction temperatures and EDX spectra of (b) BS200, (c) BS220 and (d) BS245a.

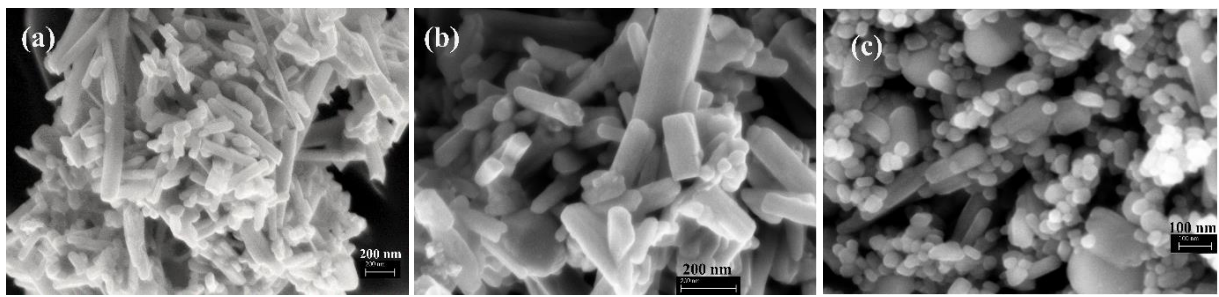

**Figure 3** FESEM images of $Bi_2S_3$-Bi nanocomposites. (a) and (b) are FESEM images of BS240 recorded with different magnifications, and (c) is that of BS245a.

applied using complete S atomic % while the remaining atomic % of Bi should be meant for Bi semimetal in each case. When we consider this in the case of chunk, Bi:S::2:3 i.e. 12.08:18.12 atomic % ratio will be for $Bi_2S_3$ and remaining 69.8% for Bi. Accordingly, they shall be 27.83:41.75 and 30.4% for nanorods and 25.72:38.58 and 35.7% for spherical NPs, respectively. They indicate that chunk-like particles contain maximum 69.8% Bi, spherical particles medium 35.7% and nanorod the lowest 30.4% Bi. These observations suggest that these two chemical phases are not of equal proportion in all the NPs but rather in varying proportions. With all the probability, considering also that size of (21.6 – 41.3 nm) of $Bi_2S_3$ are generally smaller than those (44.7 – 58.7 nm) of Bi NPs (Table S1), the latter suggestively coats larger Bi NPs forming Bi-$Bi_2S_3$ core-shell heterostructures along with independent Bi NPs. This conclusion is based also on the possible reduced melting point of Bi NPs (at the reaction temperature of 245 °C) on nanostructuring compared to its bulk value (270 °C) along with the fact that $Bi_2S_3$ started forming at relatively low temperature such as 180-200 °C,[3] at which the reaction is generally allowed to take place for nearly 2 hrs. During this process until, the reaction reaches 245 °C, S gets perhaps evaporated turning $Bi_2S_3$ NPs S-deficient and formation of Bi ions or NPs takes place and coat $Bi_2S_3$. With increasing reaction temperature ($T_{Re}$), the quantity and size of Bi NPs in the form of chunks also increases (Tables 1 and S1). This picture of presence of independent Bi NPs corroborates well the concomitant reduction in resistivity as discussed later.

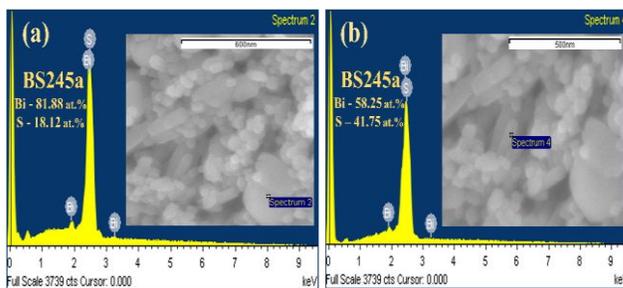

**Figure 4** EDX images of $Bi_2S_3$-Bi nanocomposites (BS245a) collected by focusing electron beam into very small portion of (a) chunk-like particle and (b) nanorod. Inset; the point of interest for EDS data (where whole electron beam is focused) is shown in its FESEM image, indicated by '**::**'.

### 3.3 X-ray Photoelectron Spectroscopy

We further characterized $Bi_2S_3$-Bi nanocomposites via XPS to confirmed valence state of constituent elements and to trace out the presence of any impurity elements. In the survey scan XPS spectrum, those peaks are observed which belong to carbon, bismuth, sulfur and oxygen only (Fig. 5 (a)), confirming the absence of any impurity elements. High resolution XPS spectra of Bi 4f, S 2s and O 1s are collected and to evaluate the accurate positions (binding energy) of peaks the curve fitting, using XPSPEAK41 software, are performed (Figs. 5 (b)–(d)). Bismuth 4f peaks are de-convoluted into two pairs of peaks, first prominent pair of peaks are centered at 158.34 eV (Bi $4f_{7/2}$) and 163.59 eV (Bi $4f_{5/2}$) with spin orbit separation of binding energies (BEs) 5.25 eV correspond to the $Bi^{3+}$ present in $Bi_2S_3$.[3,17] Second pair of peaks towards lower binding energy centered at 157 eV (Bi $4f_{7/2}$) and 162.09 eV (Bi $4f_{5/2}$) correspond to metallic $Bi^0$.[18,19] This confirms the presence of the two different valence states of bismuth, $Bi^0$ and $Bi^{3+}$, in $Bi_2S_3$-Bi composites. From above fitting one more peak centered at 161.2 eV is observed which is consistent with BE of S 2p. The BE of the S 2s peak (Fig. 5(b)) found at 225.14 eV is consistent with presence of $S^{2-}$ in $Bi_2S_3$.[3,17] The O 1s peak is observed at 531.58 eV (Fig. 5 (d)) due to adsorption of reaction by-product on the surface of the NPs[3,7] because normally metal-oxide peaks exhibit at lower BE.[19]

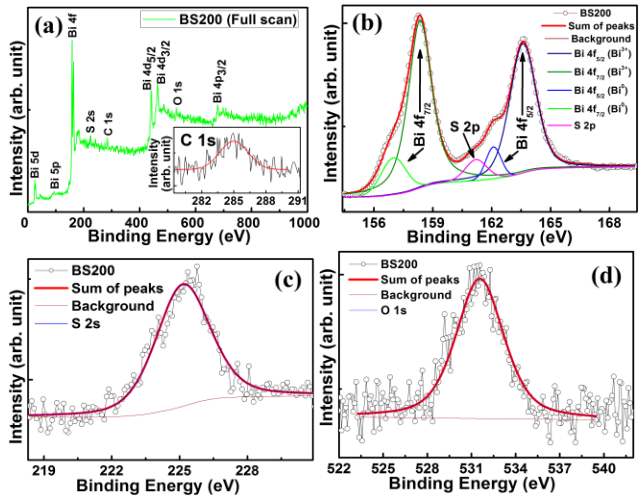

**Figure 5** (a) Survey scan XPS spectrum of BS200 and its high resolution spectra for (b) Bi 4f, (c) S 2s and (d) O 1s with fitting curves. Inset of (a) is the high resolution spectrum of C 1s. Here red solid line is the summation of all individual peaks (solid lines with different colors).

### 3.4 Zeta potential study

As above discussion reflects that Bi covers the surface of $Bi_2S_3$ NRs, its surface charge study can give very good assessment for the formation of core-shell-like structures. For this, zeta potential (ζ) measurements of the composites dispersed in deionized (DI) water at 27 °C are performed (Fig. 6 (a)). We note here that in a typical ζ measurement, NPs in a colloidal solution start moving toward the electrodes with particular mobility, depending on their surface charge and size, under a constant electric field (85 V/cm).[3,20] The mobility obtained for BS200, BS240 and BS245b nanocomposites in DI water are $1.333 \times 10^{-4}$, $7.678 \times 10^{-5}$ and $-7.524 \times 10^{-5}$ $cm^2/Vs$ and ζ values are 17.12 mV, 9.85 mV and -9.65 mV, respectively. However, the ζ peak of Bi NPs is centered at -19.06 mV. This trend on ζ can be directly correlated with that of Bi content of these nanocomposites (Table 1). In other words, as Bi content increases, positive ζ value gradually decreases and turns negative, the highest magnitude being for the pure Bi semimetal at 19.06 mV. This suggests that, most likely, $Bi_2S_3$ NRs is coated with a thin layer of Bi in addition to the independent Bi NPs. Their ζ values less than ±30 mV indicate their unstable nature in DI water and are proportional to surface charge, and smaller particles exhibit higher surface charge.[20] This conclusion is in well-agreement with crystallite size calculation (Table S1).

Figure 6 (b) shows the hydrodynamic diameter of the nanocomposites at 27 °C in de-ionized water with pH 7. The average hydrodynamic diameter ($D_{av}$) of 160 nm, 168 nm, and 252 nm for BS200, BS240 and BS245b, respectively corroborates well with increase in crystallite size as found in XRD (Table S1). Bi NPs show much higher diameter (268 nm) than others due to agglomeration. The hydrodynamic diameter (say, 160 nm) is much larger normally compared to that of XRD (22 nm for BS200) due to their fundamental difference in the detection method.[3]

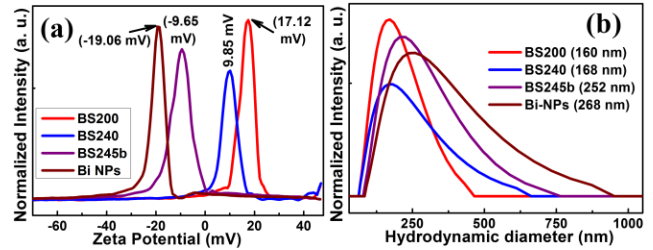

**Figure 6** (a) Zeta potential and (b) hydrodynamic diameter of BS200, BS240, BS245b and Bi NPs in deionized water at 27 °C.

### 3.5 Raman Spectroscopy

Raman spectroscopy is a powerful probe to confirm the presence of different constituent phases with different crystal structures in composites. Raman spectroscopy measurements have therefore been performed on compressed pellets of the nanocomposites. At the outset, since structure and composition of semimetal or semiconductor may be modified due to high power laser irradiations on the composite,[21] the Raman spectra are collected for BS200 and Bi NPs at different laser powers (Fig. 7 (a,b)). This problem is apparent when an Ar ion laser (λ = 473 nm) with intensity $I_0$ (100%, 19 W) is directly focused on BS200 with 99.4% $Bi_2S_3$, the distinct characteristic modes of $Bi_2S_3$ are merged with each other due to localized laser heating and producing an extra mode at 149 $cm^{-1}$ (Fig. 7 (a)). The intensity of extra peak gets suppressed and characteristic peaks are relatively resolved for applied laser power of $I_0/2$, $I_0/4$ and $I_0/10$ (Fig. 7 (a)). In Bi NPs also, clear emergence of two extra peaks (at 125 and 311 $cm^{-1}$) due to high laser power ($I_0$ and $I_0/2$) induced formation of β-$Bi_2O_3$ is seen.[21] In low laser power ($I_0/10$) only, first order Raman modes of Bi are observed, and peaks corresponding to second-order harmonic (at 185 $cm^{-1}$) of Bi[22] and β-$Bi_2O_3$ get suppressed (Fig. 7 (b)). This led us to perform Raman spectroscopic measurements with $I_0/4$ laser power for all $Bi_2S_3$-Bi nanocomposites.

In Raman spectra of $Bi_2S_3$-Bi nanocomposites (Fig. 7 (c)), nine obvious modes at 278 $cm^{-1}$, 264 $cm^{-1}$, 238 $cm^{-1}$, 188 $cm^{-1}$, 169 $cm^{-1}$, 101 $cm^{-1}$, 85 $cm^{-1}$, 74 $cm^{-1}$ and 56.5 $cm^{-1}$ are observed due to co-existence of heterostructures of orthorhombic $Bi_2S_3$ and rhombohedral Bi. These peaks are identified as $A_g$ (56.5 $cm^{-1}$, 101 $cm^{-1}$, 188 $cm^{-1}$, 238 $cm^{-1}$) and $B_{1g}$ (169 $cm^{-1}$, 264 $cm^{-1}$, 278 $cm^{-1}$) modes of $Bi_2S_3$, whereas $E_g$ (74 $cm^{-1}$) and $A_{1g}$ (98 $cm^{-1}$) modes of Bi. Interestingly, in this, intensity of an additional peak at 98 $cm^{-1}$ grows up with increasing Bi fraction (Fig. 7 (c)) and fully grown as single peak in Bi NPs (Fig. 7 (c)-(d)). Multi-peak Lorentzian fitting of all the samples are performed (Figs. 7 (d) and S3 (a-f)) and obtained peak positions with full width at half maxima (FWHM) are presented in Table S3. The FWHM of all modes of nanocomposites not changing systematically with varying Bi content may be due to the random variation of particle sizes of the components in the composites while Bi content increase seems to be systematic. But for BS200 (with 99.4% $Bi_2S_3$), the FWHM of all modes ($A_g$, $B_{1g}$ and $E_g$) are larger than those of other composites with larger Bi content. This is however consistent with smaller crystallite size samples (Table S1). The peak positions of $B_{1g}$ (264 cm$^{-1}$) and $A_g$ (101 and 188 cm$^{-1}$) modes of composites are more blue-shifted than that of pure $Bi_2S_3$[3,23] that may be due to stress-induced formation of $Bi_2S_3$-Bi core-shell-like structures or Bi-coated $Bi_2S_3$ NPs.

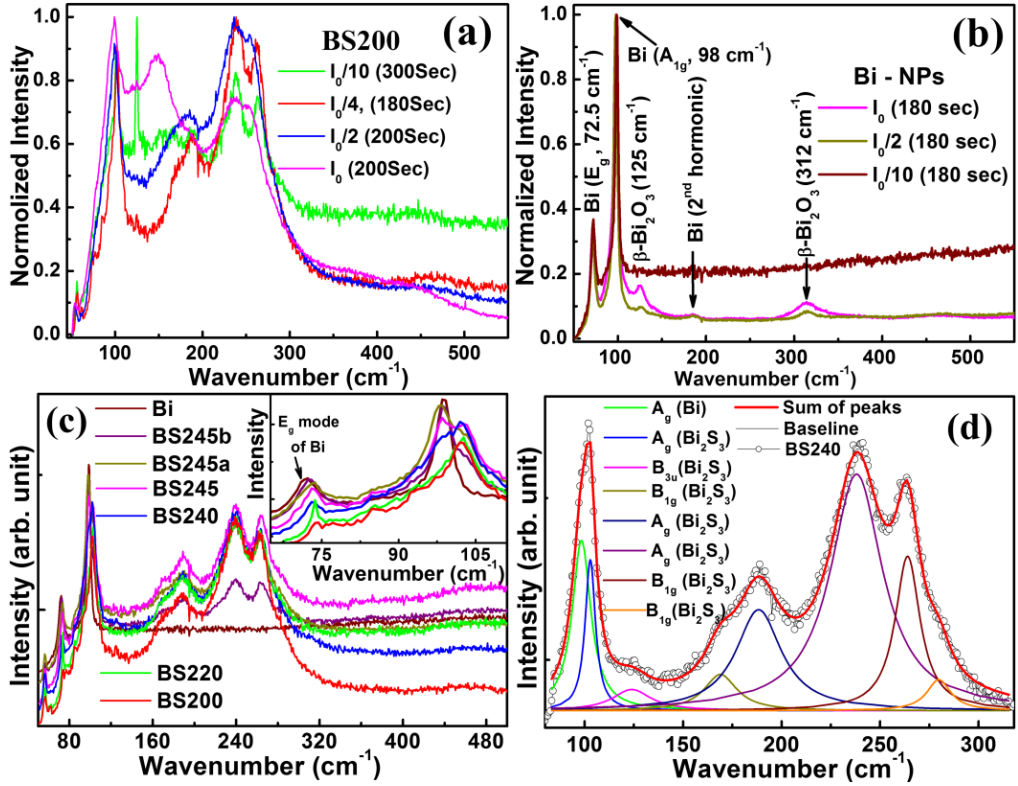

**Figure 7** Raman spectra of (a) BS200 and (b) Bi NPs collected using different laser power. (c) Raman spectra of $Bi_2S_3$-Bi nanocomposites synthesized with different reaction temperature and time. Inset shows the enlarge view of low frequency region of the spectra. (d) Multi-peak Lorentzian fitting of BS240 (black open circle) where summation of all individual Lorentzian peaks (solid line with different colors) are represented by red solid line.

Two obvious modes as $A_{1g}$ (98 cm$^{-1}$) and $E_g$ (72 cm$^{-1}$) of Bi NPs (Fig. 7 (b,c)) with slight blue-shifting from that of its bulk ($A_{1g}$ at 96.8 cm$^{-1}$ and $E_g$ at 69.2 cm$^{-1}$)[24] are consistent with earlier reports.[18,22,24,25] The FWHM of $A_{1g}$ of Bi NPs systematically decreases with increasing Bi fraction in these composites due to increase in its crystallite size with increasing $T_{Re}$/ Bi-content (Table S1 and S3). The $E_g$ optical mode (Fig. 7 (c), inset) of Bi shifts towards higher frequency side with reduction of Bi shell thickness/ crystallite size in the composites (Table S1) in agreement with earlier works.[24,25]

### 3.6 Electrical transport properties

For electrical transport properties measurements, dried powder of all the samples were consolidated in pellet forms by applying ~900 MPa pressure at 27 °C. The electrical resistivity (ρ) of BS200 is very high, not shown here, because it contained nearly 99.4 % phase fraction of $Bi_2S_3$ consistent with earlier report.[3] For reduction of ρ of $Bi_2S_3$, several efforts have therefore been already made[3,12] with the suggestion that ρ can be controlled with the creation of sulfur vacancies using vacuum annealing, hot press,[3] hot isostatic press and SP sintering[4] with however less success. We discovered a new way to reduce it by the formation of $Bi_2S_3$-Bi composites. This is possible by just increasing $T_{Re}$ and time, which increases the volatilization of sulfur that increases their vacancies in $Bi_2S_3$ and at their grain boundaries. The latter seemingly turn into nucleation centers for Bi NPs. Hence, second phase Bi comes into picture and its fraction increases in the composites with increasing $T_{Re}$ (Fig.1 and Table 1). With increase in Bi content in the composites, a systematic reduction in electrical resistivity is observed (Figs. 8 (a) and S4 (a)).

The ρ of BS220 is measured using two-probe technique which has slightly higher conductivity due to higher charge density, n (σ ∝ n) than that of BS200 due to formation of sulfur vacancies and Bi NPs. This reduction in ρ with increasing Bi content is systematically followed in other samples prepared with increasing $T_{Re}$. Therefore, ρ measurements are possible using four-probe technique for BS240, BS245, BS245a BS245b and Bi NPs (Figs. 8 (a) and S4 (a)). In overall, electrical resistivity decreases a whopping seven orders of magnitudes (~10$^7$) with increase in reaction temperature and time. This dramatically rapid reduction of ρ of highly resistive $Bi_2S_3$ is attributed to the formation of large number of sulfur vacancies at the grain boundaries region with increasing $T_{Re}$ which leads to the formation of Bi shell or Bi NPs in turn. Initially, whole electrical conduction takes place through sulfur deficient $Bi_2S_3$, such as in the case of BS240, and causes an exponentially increasing ρ with decreasing temperature (Fig. S4 (a)). Bismuth is a semimetal with higher σ (∝ n) than that of sulfur-deficient $Bi_2S_3$. Therefore, when Bi content increases sufficiently high in the composites, current percolation perhaps started through connected Bi-shell/ Bi NPs. That is, the total charge carrier density of the composites would increase as a consequence of Bi content increase and hence their ρ gets suppressed phenomenally. The ρ of all samples show exponential decay with increasing temperature near 300 K (seen more clearly when they are plotted separately) confirming their semiconducting nature (Figs. 8 (a) and S4 (a)) and a saturation-like feature below 100 K is observed in BS245, BS245a, BS245b and Bi NPs; their electrical conductivity plots are also made (Fig. S4 (b)).

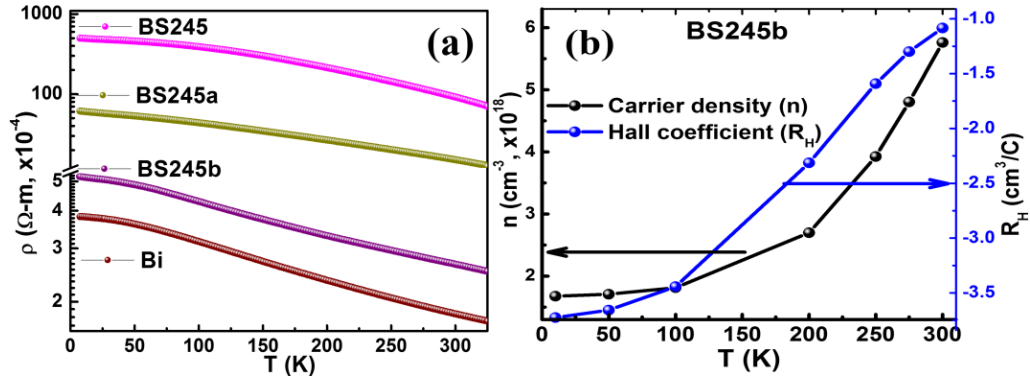

**Figure 8.** Temperature dependence of (a) electrical resistivity of $Bi_2S_3$-Bi nanocomposites synthesized at $T_{Re}$ = 245 °C for different reaction time (BS245, BS245a, BS245b) and Bi NPs, and (b) calculated Hall coefficient ($R_H$) and charge carrier density (n) of BS245b.

This scenario in ρ is due to the competition between n and mobility of charge carriers. Here, sulfur deficient $Bi_2S_3$ is n-type semiconcoductor[3] and Bi NPs with particle size near 55 nm is also expected to exhibit semiconducting nature.[12] For better understanding of transport mechanism here, we performed Hall measurements on BS245b (Fig. S5), and calculated Hall coefficient, $R_H$ and n (Fig. 8 (b)). In this, the $R_H$ of BS245b exhibits negative values only indicating electrons are its majority charge carriers and its n exponentially decreases with decreasing temperature. This confirms that the exponential decrease in ρ with increasing temperature above 100 K is due to increase in n. In ρ measurements, an applied electric current will prefer to flow via less resistive path made by internally connected Bi-shell/Bi NPs on the surface of $Bi_2S_3$ in BS245, BS245a and BS245b. In semimetal Bi, the mean free path for the carriers is nearly 1 mm[9] which is much greater than the particle size. Therefore, the mobility (μ) of charge carriers in nanostructured Bi exhibits smaller value at 300 K and will increase with decreasing of temperature[12] that outweighs the decrease in n with decreasing temperature for BS245b (Fig. 8(b)) and hence a saturation in ρ, due to ρ=1/(neμ), occurred below 100 K.

The electrical conductivity (σ = neμ) of Bi (semimetal) consist both parts conduction of electrons in conduction band ($σ_e$) and holes in valence band ($σ_h$). Therefore, Seebeck coefficient (α) of two band conduction will be

$$α = (σ_eα_e+σ_hα_h)/(σ_e+σ_h) \quad \ldots\ldots (1)$$

where $α_e$ and $α_h$ are thermopower due to electrons and holes respectively. In this way, α is directly related to σ, n and μ. Consequently, α measurement is not possible for highly resistive samples (BS200 and BS220). However, slightly less resistive sample (BS240) possesses high α due to low n and σ. α data for BS240 is not possible to collect below 136 K because of its drastically increased ρ (Fig. S4 (a,c)). Figure 9 (a) shows the α data of in the range of 5 – 325 K for $Bi_2S_3$-Bi composites. The negative sign of α in the range of 5 – 325 K implies majority charge carriers are electrons for BS245, BS245a, BS245b and Bi NPs (Fig. 9 (a)). This is well-corroborated with Hall coefficient data (Fig. 8(b)) and confirmed their n-type semiconducting nature near 300 K. The highest α achieved by resistive BS240 is -473.5 μV/K at 325 K (Fig. S4 (c)). Then, it is drastically reduced down to -163.8 μV/K for BS245 (Fig. 9 (a)) because it consists of 58 % semimetal Bi (Table 1) which has higher σ and n. In continuation, BS245a (with 65.4 % Bi and 34.6 % $Bi_2S_3$: see Table 1) shows maximum value $α_{325K}$ = -119.9 μV/K. Thereafter, $Bi_2S_3$-Bi composites with highest Bi-content BS245b (83.5% Bi and 16.5% $Bi_2S_3$) also possess significantly high α value (-97.4 μV/K at 325 K). Thus, α of BS245a is higher than BS245b mainly due to its difference in σ (or n).[1] All these composites exhibit notably high α values near 300 K that is due to low energy charge carriers scattering at the semimetal to semiconductor (SS) interface which leads to increase in average energy per charge carriers.[7,8] Finally, for 100 % Bi (Bi NPs), the highest α is -55.9 μV/K at 325 K, consistent with other report.[26]

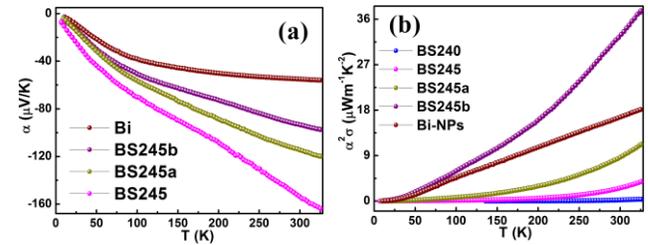

**Figure 9** Temperature dependent (a) Seebeck coefficient (α) and (b) calculated thermoelectric power factor of $Bi_2S_3$-Bi nanocomposites with Bi content prepared at different reaction parameters (temperature and time).

These results may be understood using two band model of Bi semimetal, wherein the presence of two types of charge carriers in conduction is considered.[11, 26-29] Typically, with application

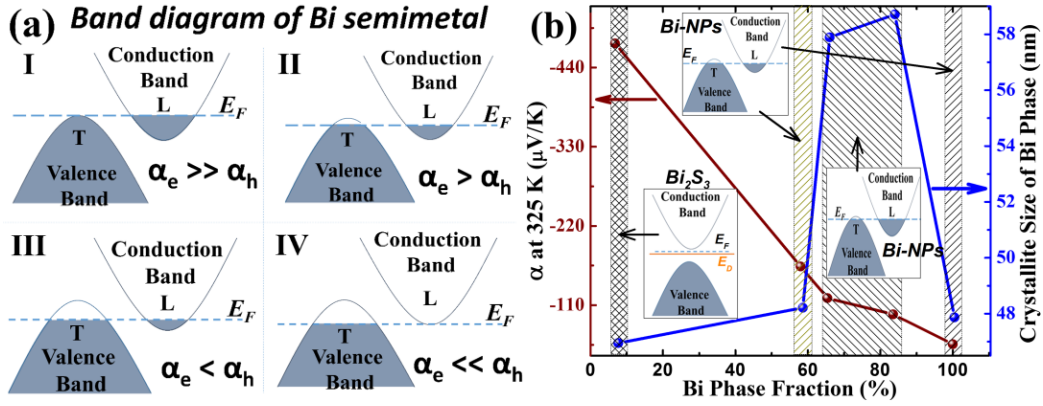

**Figure 10** (a) Schematics for competition between thermopower components due to electrons ($\alpha_e$) and holes ($\alpha_h$) corresponding to different positions of Fermi level in bismuth semimetal and (b) $\alpha$ at 325 K and calculated crystallite size versus Bi-phase fraction obtained from Rietveld refinements of $Bi_2S_3$-Bi nanocomposites with typical band diagrams.

of temperature difference across the nanocomposite, some electrons and holes start thermal diffusion in opposite directions which produced $\alpha$ with opposite sign. For semimetal Bi, when Fermi level ($E_F$) lies at the top of valence band (Case-I in Fig. 10 (a)), then transport will be driven by the electrons and exhibits only negative $\alpha$. However, when $E_F$ lies just at the bottom of conduction band (Case-IV in Fig. 10 (a)), then whole transport will be driven by holes thereby $\alpha$ exhibits positive value only. If $E_F$ lies in between above two extremes (like in Cases II and III, Fig. 10 (a)), then both types of charge carriers can participate in conduction and can exhibit positive $\alpha$ ($\alpha_h$) for holes or negative $\alpha$ ($\alpha_e$) for electrons. In this way, the sign of $\alpha$ reflects the sign of majority charge carriers. In the investigated composites, $\alpha$ values are negative in the entire temperature range indicating that Fermi level lies deep in conduction band (like Case-II). A band theory based on a systematic reduction of band overlap with reducing particle size of Bi has been assigned here, as in ref. 10, as depicted in Fig. 10 (b) for the present system.

In the case of Bi semimetal, the dominance of particular types of charge carries is highly dependent on the particle size and shape.[11,12,27] The effective mass of electron ($m_n^* = 0.0204\ m_e$) is lighter than that of hole ($m_p^* = 0.69\ m_e$)[30] and mobility ($\mu$) is inversely proportional to mass. Thus, since the mobility of electrons ($\mu_e$) is higher than the mobility of holes ($\mu_h$) in bulk Bi, it produces smaller $\alpha_e$ than $\alpha_h$ and hence the total $\alpha$ value is higher.[1,27-29] For example, in bulk Bi semimetal, theoretical calculations give $\alpha_e = -486$ μV/K and $\alpha_h = 669$ μV/K which produces relatively low positive total $\alpha$ at 300 K.[29] When grain size is reduced, then $\mu_e$ decreases relatively faster than $\mu_h$ due to enhanced grain boundary scattering of electrons which leads to decrease in $|\alpha_h|-|\alpha_e|$. Consequently, low dimensional materials like NPs, nanorods, nanowires[12,26] and thin film[11] exhibit lower $\alpha = \alpha_h + \alpha_e$ than that of the bulk. Therefore, Bi NPs with average L of 47.8 nm exhibit the highest value -55.9 μV/K at 325 K that is consistent with literature.[12,26] In BS245b, L of Bi phase is 58.7 nm which is greater than that of pure Bi NPs (47.8 nm) thereby magnitude of $\alpha$ of BS245b is greater than (-97.4 μV/K at 325 K) that of pure Bi NPs (-55.9 μV/K at 325 K).

As the Fermi surface of bulk Bi in the Brillouin zone consists of three highly elongated electron pockets at the L-point[11,12] and the band gap at T-points ($E_{gL}$) increases with decreasing Bi wire diameter.[10] Herein, the sign of thermopower, combined with Hall coefficient, analysis confirms that majority carriers in these composites are electrons. Thus, to check the effect of particle size reduction on $E_{gL}$ of semimetal Bi for the Bi covered $Bi_2S_3$ nanocomposites, we performed two-band model (TBM) fitting for temperature (T)-dependent $\alpha$ data[31,32] using a simplified expression of Eq. (1) for the semimetals as $\alpha = aT + (bE_{gL}/2k_B + cT)\exp(-E_{gL}/2k_BT)$ in which a, b and c are fitting parameters. The TBM fitting curves of $\alpha$ data are shown in Fig. S6 and obtained parameters are presented in Table S4. Obtained $E_{gL}$ is 32.5 meV, 11.1 meV, 9.6 meV and 11.9 meV for BS245, BS245a, BS245b and Bi NPs respectively, well in accordance with the crystallite size increment of Bi phase in the composites (Table S1), consistent well with earlier report.[10]

Now, the thermoelectric power factors (PFs) are calculated for BS240, BS245, BS245a, BS245b and Bi NPs (Fig. 9(b)). The suppression of $\rho$ by just optimization of reaction parameters through formation $Bi_2S_3$-Bi nanocomposites are quite interesting results in the direction of improvement of TE performance of resistive $Bi_2S_3$. While the highest $\alpha$ value of -473.5 μV/K is achieved for BS240 at 325 K, but due to its correspondingly high $\rho$ value, the TE power factor is very low (0.36 μW/mK$^2$ at 325 K). The highest PF is 37.67 μWm$^{-1}$K$^{-2}$ at 325 K for BS245b, whereas other samples (BS245 and BS245a) with moderate $\rho$ achieved reasonably high PF (Table S5). This is nearly 3 orders of magnitude greater than that of about 0.03 μW/mK$^2$ at 300 K of the pure $Bi_2S_3$ sample, hot pressed at 480 °C.[3] Ge et al.[4] achieved the highest ZT of 0.36 at 623 K for their SP sintered (at 673 K) core-shell $Bi_2S_3$@Bi 1h sample that exhibited a PF of ~29 μWm$^{-1}$K$^{-2}$ at 325 K, which is 23 % less than that of the presently studied BS245b (37.67 μWm$^{-1}$K$^{-2}$). This higher PF, for nanostructured $Bi_2S_3$-Bi nanocomposites without any sintering steps, which do not allow Bi NPs to expel out turning into bulk Bi with probable considerably reduced thermal conductivity, is brilliantly substantial. Lower $\kappa$ is expected for these composites due to scattering of phonons from the $Bi_2S_3$-Bi interfaces and point defects in composites, which would lead to greatly improved thermoelectric performances. This work will therefore be a very pioneering one for development of toxic- and rare earth element-free TE materials near room temperature.

## 4. Conclusion

Novel $Bi_2S_3$-Bi nanocomposites with different Bi contents are synthesized in one step polyol method. The existence of double phase is confirmed by Rietveld refinement analysis whereby change in phase fraction is successfully obtained. Zeta potential study shows a systematic reduction and sign reversal in surface charge with increasing Bi fraction suggesting the formation of Bi covered $Bi_2S_3$ nanoparticles. A colossal reduction in electrical resistivity ρ of nearly seven orders of magnitudes (~$10^7$) is observed with increasing reaction temperature and time which came out to be a very significant success in the direction towards reducing high ρ values of $Bi_2S_3$. Most interestingly, thermoelectric (TE) power factor at 300 K of highest Bi content nanocomposite pellet, pressurized at just 300 K, is 3 orders of magnitude greater than that of hot-pressed $Bi_2S_3$ and even 23% better than that of spark plasma-sintered core-shell $Bi_2S_3$@Bi sample reported earlier, showing greatly improved TE performances (due to the probable greatly reduced thermal conductivity) in the present method of synthesizing these samples. Hence, present work will serve as a pioneering approach for development of toxic- and rare earth element-free TE materials near room temperature.

## ASSOCIATED CONTENT

Electronic supporting information includes: Figures of Rietveld refinements of XRD patterns, EDX, Lorentzian fitting of Raman spectra, $ρ_{xy}$ versus magnetic field scan graph and temperature dependent S, ρ and σ data of nanocomposites. Also includes some Tables for calculated crystallite size and mass density, atomic percentages obtained from EDX, FWHM and positions of Raman modes, obtained fitting parameters from α data, and values of S, σ and $S^2σ$ at 325 K for nanocomposites.

## AUTHOR INFORMATION

*Dr. Gunadhor Singh Okram,
Email ID – okramgs@gmail.com, okram@csr.res.in
ORCID iD -  0000-0002-0060-8556

## ACKNOWLEDGMENT

Authors are gratefully acknowledge to Mukul Gupta & Layanta Behera and D. M. Phase & V. K. Ahire for XRD and EDX data from UGC-DAE Consortium for Scientific Research, Indore.

SYNOPSIS TOC "Thermopower of Bi$_2$S$_3$-Bi core-shell nanocomposites, synthesized with different Bi content/shell thickness by polyol method at different reaction temperature and time, with typical band diagram for corresponding samples."

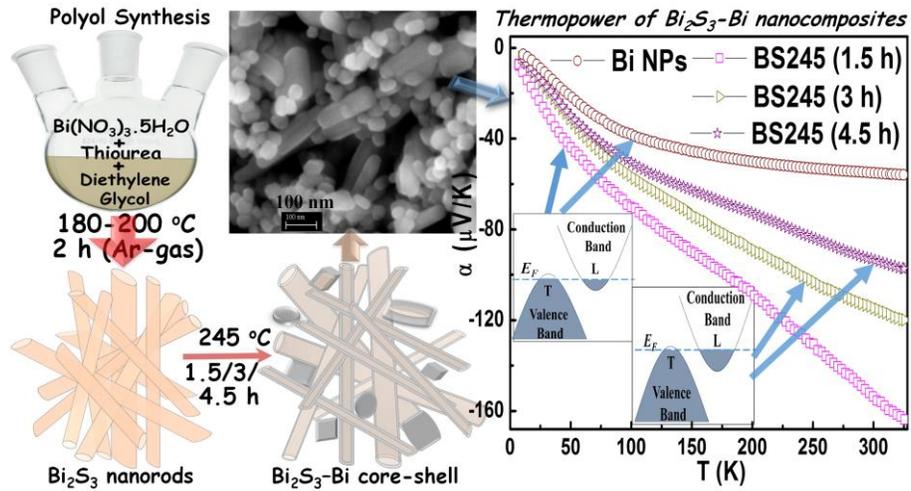